\documentclass[sigconf]{acmart}

\usepackage{booktabs} 
\usepackage{balance}
\usepackage{filecontents}
\usepackage{xurl}

\setcopyright{rightsretained}
\usepackage{tabularx}

\begin{document}
\balance
\title[Twitter TrollHunter/Evader]{TrollHunter [Evader]: Automated Detection [Evasion] of Twitter Trolls During the COVID-19 Pandemic}

\author{Peter Jachim}
\affiliation{%
  \institution{DePaul University}
  \streetaddress{243 S Wabash Ave}
  \city{Chicago}
  \state{IL}
  \postcode{60604}
}
\email{pjachim@depaul.edu}

\author{Filipo Sharevski}
\affiliation{%
  \institution{DePaul University}
  \streetaddress{243 S Wabash Ave}
  \city{Chicago}
  \state{IL}
  \postcode{60604}
}
\email{fsharevs@cdm.depaul.edu}

\author{Paige Treebridge}
\affiliation{%
  \institution{DePaul University}
  \streetaddress{243 S Wabash Ave}
  \city{Chicago}
  \state{IL}
  \postcode{60604}
}
\email{ptreebri@cdm.depaul.edu}

\renewcommand{\shortauthors}{P. Jachim, F. Sharevski, P. Treebridge}

\begin{abstract}
This paper presents \textit{TrollHunter}, an automated reasoning mechanism we used to hunt for trolls on Twitter during the COVID-19 pandemic in 2020. Trolls, poised to disrupt the online discourse and spread disinformation, quickly seized the absence of a credible response to COVID-19 and created a COVID-19 infodemic by promulgating dubious content on Twitter. To counter the COVID-19 infodemic, the \textit{TrollHunter} leverages a unique linguistic analysis of a multi-dimensional set of Twitter content features to detect whether or not a tweet was meant to troll. \textit{TrollHunter} achieved 98.5\% accuracy, 75.4\% precision and 69.8\% recall over a dataset of 1.3 million tweets. Without a final resolution of the pandemic in sight, it is unlikely that the trolls will go away, although they might be forced to evade automated hunting. To explore the plausibility of this strategy, we developed and tested an adversarial machine learning mechanism called \textit{TrollHunter-Evader}. \textit{TrollHunter-Evader} employs a Test Time Evasion (TTE) approach in a combination with a Markov chain-based mechanism to recycle originally trolling tweets. The recycled tweets were able to achieve a remarkable 40\% decrease in the \textit{TrollHunter's} ability to correctly identify trolling tweets. Because the COVID-19 infodemic could have a harmful impact on the COVID-19 pandemic, we provide an elaborate discussion about the implications of employing adversarial machine learning to evade Twitter troll hunts.
\end{abstract}



\maketitle



\section{Introduction}
Pandemics are sustained emergencies that result in societal crisis due to an absence of optimal response in containing an unprecedented threat to  public health. The emergencies during a pandemic, at least in recent history, are exacerbated by a lack of trust in information from authorities because state-sponsored actors fill the void with disinformation and rumours \cite{Kassam}. In the 1980s, for example, the KGB initiated an information warfare campaign called ``Operation Infektion'' to spread the rumour that HIV/AIDS was a misfired American biological weapon in order to undermine the United States' credibility during the Cold War \cite{Boghardt}. Published and amplified through various foreign media outlets, the campaign achieved a relative success in manipulating  public opinion and shifting attention away from Russian biological weapon programs ~\cite{Grimes}.  

The COVID-19 pandemic in 2020 brought another opportunity for public opinion manipulation \cite{who}. This time, because many people were forced to stay home, they went online and actively followed news and participated in public discourse on various social media platforms, where they encountered diverse disinformation and rumours spread around the globe instantaneously \cite{Frenkel}. The public discourse, as such, is conducive to disinformation and rumours because it lacks a broad editorial inspection, users 
are free to contribute a wide variety of content, interact with anybody and form various types of groups, agendas and discussion topics at no cost \cite{Kirman}. The online discourse allows users to engage in antisocial and often malicious behaviour known as \textit{trolling} ~\cite{Coles}. 

A \textit{troll} is a user who constructs the identity of sincerely wishing to be part of the group in question, including professing, or conveying pseudo-sincere intentions, but whose real intention(s) is/are to cause disruption and/or to trigger or exacerbate conflict in discourse \cite{Hardaker1}. Thus, \textit{trolling} refers to ``a specific type of malicious online behaviour, intended to disrupt interactions and the general online discourse, aggravate conversational partners and lure them into fruitless argumentation'' \cite{Coles}. For some time, trolling was antisocial behaviour characteristic of the gaming communities and fringe discussion forums like ``4chan'' \cite{Samory, Kirman}. Trolling quickly spread on social media where the trolls were not simply ``amusing themselves by upsetting other users,'' but intentionally disseminating disinformation as part of a state-sponsored effort to manipulate public opinion about political candidates, public health issues like vaccination and social justice issues \cite{Zannettou, Llewellyn, Broniatowski, Stewart}. The definition of \textit{trolling} extends to encompass ``users who exhibit a clear intent to deceive or create conflict with the goal to manipulate the public opinion on a polarised topic and cause distrust in the socio-political system'' \cite{Badawy}. The methods used for trolling evolved from using  offensive language to dissemination of disinformation, rumours and fake news \cite{Benkler}. 

From sporadic activity by individual users, trolling became a orchestrated effort of groups of users or \textit{troll farms} that coordinate the dissemination of trolling content on social media platforms, often using \textit{social bots} to amplify by creating misperceptions of consensus on a polarised topic \cite{Cresci}. Trolling took a form of political information operations, sponsored by nation-states, aiming to disrupt a constructive process of political deliberation on Twitter \cite{Twitter}. The detection and eradication of trolling developed into a serious problem for social media platforms because the state-sponsored troll farms are persistent in their efforts, using both fake and bot accounts and have a wide array of polarised topics to choose from to create trolling content and manipulate public opinion. Usually, social media platforms rely on moderators for banning/muting trolling users and flagging/deleting trolling content, but this kind of manual solution has some major drawbacks, including a delay of actions, subjectivity of judgment and scalability \cite{Ortega, Fornacciari}. The need for automated trolling detection thus drew the attention of the research community yielding several popular approaches including linguistic and sentiment analysis \cite{Seah, Ghanem, Capistrano, Fornacciari}, metadata analysis \cite{Mihaylov, Fornacciari} and social network analysis \cite{Kumar, Riquelme}. 

The lack of immunization and global political coordination to handle the 2020 COVID-19 provided a fertile ground for trolls to manipulate public opinion on social media, especially on Twitter \cite{Madrigal}. From conspiracy theories about the provenance of the virus as a biological weapon, its relation to 5G cellular network technology, rumours about how people suffer and die from COVID-19, to disinformation about the efficacy of alternative drugs, the trolling activity on Twitter became an ``infodemic'' that spreads faster and more easily than COVID-19 itself, threatening both online discourse and public health \cite{who}, \cite{Starbird2020}. In order to proactively help with the detection of this COVID-19 infodemic, we developed an automated reasoning mechanism called \textit{TrollHunter} that leverages machine learning and linguistic analysis to hunt for trolling content on Twitter. We trained and tested the \textit{TrollHunter} using a dataset of 1.3 million tweets collected in a period of January-March 2020 and achieved an accuracy of 98.5\%, precision of 75.4\% and recall of 69.8\%. The first part of our paper describes the design of \textit{TrollHunter} and elaborates on its performance in detail.   

The COVID-19 pandemic is unlikely to loosen its grip until a vaccine and prevention measures are developed and implemented globally \cite{Cohen}. It is reasonable to expect that the COVID-19 infodemic is also going to persist on Twitter, with evolved methods for evading detection for trolling behaviour. One of these methods is the use of an adversarial machine learning technique called ``Test Time Evasion (TTE)'' where adversaries can gain an advantage over a deployed machine learning model for trolling detection by figuring out how the model works and changing the input sample to cause the model to misclassify a trolling tweet as a ``non-trolling'' \cite{Huang}. Employing a variant of TTE in combination with the linguistic manipulation technique known as Ambient Tactical Deception (ATD) \cite{nspw2019}, we developed and tested an automated tool called \textit{TrollHunter-Evader} to evade trolling detection. The \textit{TrollHunter-Evader} was able to reduce the \textit{TrollHunter's} performance of accurately detecting trolling tweets by 40\%. 

The \textit{TrollHunter-Evader} utilises a Markov chain to replace the words and hashtags in a trolling tweet that \textit{TrollHunter} uses to make its distinction between trolling and non-trolling tweets. The \textit{TrollHunter-Evader} employs a new evasion paradigm where an existing tweet, either ``trolling'' or ``non-trolling'' can be re-cycled and re-purposed for a persistent trolling campaign without the need to continuously develop new trolling content in the effort to evade detection. The second part of our paper describes the \textit{TrollHunter-Evader} model, the use of the ATD technique to choose target replacement words and hashtags and its performance in evading the \textit{TrollHunter}. We are aware that our work presenting a tandem of detection/evasion models for trolling detection has both practical and ethical implications. In the last part of the paper we discuss these implications and weigh the value of publicly sharing a proof-of-concept adversarial machine learning paradigm with knowledgeable researchers as an effort to proactively thwart the COVID-19 infodemic.  

\section {Related Work: Trolling Classification Models}
Detecting trolling is a complex task because anyone can post trolling content online. In a study analyzing the antecedents of individual trolling behaviour, Cheng et al. found that mood and discussion context together can explain trolling behaviour better than an individual's history of trolling \cite{Cheng}. Analysing the state-sponsor trolling linked to the Russian troll farm Internet Research Agency (IRA), one study found that trolls create a small portion of an original trolling content (e.g. posts, hashtags, memes, etc) and heavily engage in retweeting around a certain point in time of interest (e.g. the Brexit referendum) \cite{Llewellyn}. An investigation into the trolling activity around the 2016 US elections reveals different state-sponsored strategies: IRA trolls were pro-Trump while Iranian trolls were anti-Trump. Both troll farms were not consistent over time in disseminating trolling content to evade straightforward detection of their content on social media platforms \cite{Zannettou}. Another aspect of trolling detection is social media platforms' goal to allow for a high degree of participation and constructive public discourse, making them reluctant to immediately exclude users exhibiting trolling behaviour to avoid perceptions of excessive control and censorship \cite{Fornacciari}. 

The need for automated detection of trolling in social media ecosystems is evident given the threat to the integrity of the online discourse and the credibility of public opinion posed by trolls' diverse motives, forms and types. Various researchers have proposed troll detection algorithms to solve the trolling problem. Because trolling content is mostly textual in nature and comes in a form of a social media post or a comment that contains inflammatory and hostile language, one approach for trolling detection is to employ a \textit{linguistic and sentiment analysis}. A domain-adapting sentiment analysis used to detect trolls using post-level, user-level and thread-level linguistic features showed a promising performance of detecting trolls around 70\% of the time in online forums \cite{Seah}. Measuring sentiments and emotions of posts have also helped discriminate trolls on Twitter. Fornacciari et al. evaluated the “abusiveness” of a text with other metadata from trolling posts to detect Twitter trolls more than 76\% of the time \cite{Fornacciari}. Capistrano, Suare and Naval used an out-of-box sentiment analyzer called VADER \cite{VADER} in combination with lexical, syntactic and aggression analyzers and achieved 88.95\% accuracy, 86.88\% precision and 93.12\% recall when tested with the Kaggle Twitter cyber-trolls dataset \cite{Capistrano, Kaggle}. Another trolling detection algorithm analyzing the writing style of the IRA Twitter trolls looked into the emotional, morality and sentiment changes showing a 0.94 F1 score ~\cite{Ghanem}.  

Social media and online forums provide a wealth of metadata about their users that research has also utilized towards learning and detecting trolling behaviour. Augmenting the sentiment analysis with information about the publication time of the trolling posts (workday, weekend, work time, night time), Mihaylov and Nakov created two classifiers: one for detecting “sponsored trolls”, who try to manipulate a user's opinion and one for detecting classical “individual trolls”, who offend users and provoke anger \cite{Mihaylov}. Both detection algorithms achieve similar accuracy of 82\%. Kumar, Spezzano and Subrahmanian utilised the metadata in a larger social network analysis of the Slashdot Zoo platform achieving 51\% average precision when detecting trolling users \cite{Kumar}. Although not directly aimed at detecting trolls, Riquelme and Gonzalez-Cantergiani. showed an interesting approach for analysis of users' activity and influence on Twitter that can be used to discriminate between normal users, individual trolls, or state-sponsored trolls \cite{Riquelme}.

\section{TrollHunter: Automated Twitter Trolling Detection}
The approaches for trolling detection reviewed in the previous section provide a good basis for a broader automated reasoning when hunting for trolls on various online platforms. Because the troll detection feature sets, the context of the discourse, and the datasets vary greatly between different algorithms, we adapted our automated reasoning for troll hunting to utilise a specific feature set of Twitter trolling content during the COVID-19 pandemic in early 2020 (January-March 2020). This section describes the dataset creation and the design of our \textit{TrollHunter} algorithm for detecting the anti-social behaviour behind the COVID-19 infodemic exhibited both by individual and possibly state-sponsored trolls. In the context of our analysis, we were not concerned about whether a user is a troll, but instead that user is tweeting trolling content. 


For this purpose we established a broader definition of the \textit{Twitter trolling content} as \textit{content that is created with the intent to deceive and create a conflict with the goal to manipulate the public opinion about the COVID-19 pandemic and cause distrust in the socio-political system}. This Twitter trolling content could be leveraged for information operations, for example, the People's Republic of China Twitter offensive for ``cheerleading for the Chinese government, criticizing the US pandemic response, quibbling over the international perception that Taiwan’s response was superior to China’s, and attacking Guo Wengui for allegedly spreading false news on the coronavirus and `discrediting China''' \cite{Strick}. It could also be leveraged to wage an emotional attacks towards individual users, for example, exploiting their anxiety resulting from the intolerance to uncertainty surrounding the COVID-19 pandemic \cite{Mertens}. Our definition  follows Twitter's assessment rules for misleading information about the COVID-19 pandemic, considering any ``assertion of fact (not an opinion), expressed definitively, and intended to influence others’ behavior about the origin, nature, and characteristics of the virus; preventative measures, treatments/cures, and other precautions; the prevalence of viral spread, or the current state of the crisis; official health advisories, restrictions, regulations, and public-service announcements; and how vulnerable communities are affected by/responding to the pandemic.'' ~\cite{TwitterCovid}. 

The deceptive nature of this content is represented with tweets that include incorrect statements about the COVID-19 pandemic, for example, that the pandemic is engineered by humans as part of a biological warfare and fits as an element into a broader conspiracy for the political aspirations of China to dominate the world and weaken the United States. The conflicting nature of this content represents tweets that aim to provoke users to participate in a  fruitless debate, for example, that the COVID-19 virus is a hoax engineered by the Democratic Party, or that the virus is a sufficient reason for boycotting the WHO and China. The Twitter trolling content aims to cause distrust in the socio-political system by portraying a distorted image of inefficiency, secrecy, conflict and unsupported accusations of inadequate response to the COVID-19 pandemic. 

Certainly, our definition is restrictive compared to the general consideration of trolling as ``strategic provocation and harassment to cause maximum chaos'' \cite{Brunton}, which spreads across the entire web of false information, not just on Twitter and not just on the COVID-19 pandemic. We also hunt for those ``digital constituents'' that exploit the divergent understanding and use within the Twitter community, but our focus is on their product specific to the COVID-19 infodemic and not necessarily their entire spectrum of trolling or otherwise antisocial behaviour. Our approach for COVID-19 troll hunting on Twitter does not explicitly distinguish between the various types of false information (e.g. conspiracy theories, hoaxes, rumors, propaganda, fallacies, etc.), the actors behind it (e.g. bots, hidden payed posters, state-sponsored trolls, true believers, etc.) or the actors' motives (e.g. political Influence, malicious intent, profit, fun) \cite{Sirivianos}. Unlike the Twitter approach for singling out the accounts disseminating trolling content in order to further investigate their background \cite{Twitter}, we only focus on detecting the trolling content disseminated by these or individual accounts in order to capture a better representation of the COVID-19 infodemic. We are aware that a downside of this approach is that we are unable to account for the context of a twitter interaction or a history of a speaker and their past Twitter behavior. We acknowledge these limitations and the possibility to mistakenly define honest content as Twitter trolling content, which in turn, can disproportionately affect opinions that we perceive as inaccurate given our perspectives and creates a narrow narrative at the expense of the present diversity of opinions. 


\subsection{Dataset Creation}
\subsubsection{Data Collection}
In our study, we decided not to work on a Twitter dataset with already identified trolling users and trolling tweets (e.g. the IRA troll dataset used by Ghanem, Buscaldi and Rosso \cite{Ghanem} or the DataTurk's trolling dataset \cite{Kaggle}) nor to use Fornacciari et al's approach to rely on user reports of twitter trolls/posts  \cite{Fornacciari}. Instead, we utilised the Twitter API to collect a general set of tweets related to the COVID-19 pandemic that contained one of the following keywords: ``covid19,'' ``coronavirus,'' ``corona virus,'' ``coronavirus pandemic,'' or ``corona outbreak,'' as well as tweets that mentioned either the ``@CDC,'' or ``@WHO'' twitter accounts. We chose those searches to limit our analysis to tweets where the person posting the tweet chose to deliberately create an association with either COVID-19 or one of the larger entities fighting the illness when we started collecting tweets. We collected the tweets in real-time using a python script that ran overnight on a designated server and put the tweets into Comma Separated Value (CSVs) files which we were able to use for the labelling and could be easily imported into our model. We decided to use this strategy to help ensure that the tweet content would not be affected should a tweet be deleted. In total, we collected 1.3 million tweets. From those, two researchers manually labelled 25,000 tweets to identify potentially trolling content. The labelling took roughly 3 weeks. The inter-coder reliability was acceptable (Cohen's $\kappa=0.83$). 

\subsubsection{Data Labelling}
To label a tweet as a ``trolling'' one, we used a combined approach of identifying the tweets or comment based on the distinct individual \cite{Samory} and state-sponsored \cite{Zannettou} trolling features. We applied a topic modeling technique on our dataset to label tweets as ``trolling'' based on several disinformation campaigns and rumours already circulating as part of the COVID-19 infodemic when we collected the tweets. We manually labelled the randomly selected 25,000 tweets based on the following criteria: (1) trolling hashtags; (2) trolling topics; (3) sentiment; and (4) user behaviour. We labeled the tweets directly in the CSVs containing the tweets from the overall dataset. This means that the tweets were labelled in a slightly different context than the context in which they were initially tweeted. This was beneficial because it forced us to focus primarily on the content in each of the tweets rather than being impacted by the profile picture, or any pictures which might be included in the tweet.  

For the trolling tweets and trolling topics, we looked for content that we suspected was intended to sow disinformation or unverified rumours, based on Benkler, Faris and Roberts' broader definition of the associated behaviour \cite{Benkler}. We identified the topics based on those catalogued by the World Health Organization \cite{who} under their ``mythbusters'' section and under the Twitter policy of misleading information about the COVID-19 pandemic \cite{TwitterCovid}: there are currently no drugs licensed for the treatment or prevention of COVID-19, no folk medicine has proven effective, the virus is not human engineered, and the virus has nothing to do with 5G. We also included political themes that have been characteristic for the state-sponsored trolls such as political hoaxes, partisan blaming, international politics, and Trump \cite{Zannettou}. In addition to the above thematic identification, we looked for tweets that had a strong negative sentiment, given that such tweets more likely to have been created with the intention of creating discord \cite{Lokk}.

While we invested a large amount of effort to be as impartial as we could, it is important to note our subjective and conservative labelling of ``trolling content'' may be inaccurate because we may have misinterpreted the intentions of the tweet, held a slightly different or broader definition of trolling, or failed to detect trolling altogether \cite{Hardaker}, \cite{Samory}. We certainly brought a level of ``labeling bias'' stemming from the inability to stay completely impartial and apolitical given that we, as people, are also affected by the COVID-19 pandemic and its political contextualisation \cite{Jiang}. An early study, for example, showed a relationship between people's political attitudes and the sensitivity to the COVID-19 threat, the political framing of the pandemic, and the perception of the related false information \cite{Calvillo}. The manual labelling may have affected our judgment to disproportionately mislabel select perspectives as trolling or non-trolling, given that the COVID-19 infodemic was evolving during the time of our dataset creation. We are aware that our labelling approach poses a limitation to the automated trolling detection and welcome revisions of the dataset creation criteria to better capture the COVID-19 infodemic content.

\subsection{Feature Engineering}
This section shows the steps we performed to prepare the dataset for machine-learning processing. After we created the dataset, we enhanced the text in a number of ways to help increase its ability to help us identify trolling content including casting a wider net for detecting trolling content (hashtags, tropes, sentiment analysis and user behaviour). In addition to common techniques for creating numeric vectors from text which we used in the classification model, we noted that specific phrases and hashtags were being recycled and mimicked by a large number of people trying to advance a particular narrative. While these hashtags, phrases and thematic approaches to sharing one's voice could be used in non-trolling contexts, in conjunction with the other features, including the sentiment, ratio of capitalised to non-capitalised text, the other words and hashtags, and the machine learning approach, it is considerably less likely that one of these phrases we identified would be solely responsible for causing our algorithm to mis-classify a tweet. 

\subsubsection{Trolling Hashtags}
After manually labelling the dataset, we isolated several of the following hashtags that corresponded to the COVID-19 infodemic as defined by the WHO \cite{who}. The specific hashtags that we identified included:

\begin{itemize} 
    \item \textbf{People Died}: tweets using the strings ``liedpeopledied,'' and ``hideandpeopledied.'' Examples of these hashtags include \#ChinaLiedPeopleDied, \#WHOliedPeopleDied and \#ChinaHideAndPeopleDied.
    \item \textbf{Sinophobic}: tweets attempting to associate COVID-19 with China. The search strings we used to identify these tweets were ``china,'' ``ccp,'' ``wuhan,'' ``chinese,'' ``xijinping,'' ``xi,'' and ``wetmarkets.'' A few of the hashtags that we identified included: \#ChinaVirus, \#CCPvirus, \#ShameOnChina, \#ChinesePlague and \#MakeChinaPay.
    \item \textbf{Iran}: tweets that include a hashtag which contains the word ``Iran.'' This included tweets like: \#liftsanctionsoniran and \#helpirancoronawhom.
    \item \textbf{Trump}: hashtags mentioning Trump, based on the use of the string ``trump.'' Hashtags included \#TrumpPandemic, \#TrumpVirus and \#TrumpLiesAboutCoronavirus. 
    \item \textbf{Rumors}: hashtags supporting information like a religious stance or misleading statistic about COVID-19:  \#DiseaseFree\_With\_TrueWorship and \#coronavirusstats.
    \item \textbf{Hoax}: hashtags including the word ``hoax.'' Actual hashtags include \#CoronaVirusHoax, \#DemocraticHoax and \#CoronaVaccineHoax.
    \item \textbf{General Negativity}: hashtags promoting a negative viewpoint. A few of the search terms that we used included ``\#waronhumanity,'' and ``\#enemywithin.''
    \item \textbf{Policy} included any hashtags that reference specific policy changes. The search terms used included ``liftsanction,'' ``banchina,'' ``boycottchina,'' and ``defenseproductionact.'' Actual hashtags that we identified included \#DefenseProductionActNow and \#BoycottChina.
    \item \textbf{Right-Leaning}: hashtags referencing left-leaning American politics. Search terms included ``democrats,'' ``pelosi,'' and ``communist.'' Actual hashtags included: \#PelosiHatesAmericans and \#DemocraticHoax.
\end{itemize}
We constructed a feature to identify whether one of the hashtags or one of the patterns outlined in this list was identified.

\subsubsection{Trolling Tropes}
While labelling the data, beyond the general trolling topics provided by WHO \cite{who}, we noted that in the tweets which we labelled as being trolling tweets, we identified a few very specific tropes in our dataset. These tropes were super-specific strings that were commonly used to advance trolling topics \cite{who}, specific tropes included:

\begin{itemize}
    \item \textbf{Fake Cures}: tweets referencing one of the fake cures being presented. This was flagged if a tweet contained either ``chloroquine,'' or ``paracetamol,'' in addition to ``efficacy,'' or the stems ``effect'' (which goes into ``effective'' and ``positive effect'') or the stem ``medic'' (which was commonly used in ``medical professionals,'' or with the word ``medicine''). 
    \item \textbf{Russian Scientist} identified whether the tweet mentioned the words ``Russian scientist''.
    \item \textbf{Culpable Death}: tweets attributing someone's death to someone else. These were identified by tweets that mentioned the words ``culpable'' and ``death.''
    \item \textbf{Bat Eaters}: tweets mentioning the stereotype that Chinese people eat bats. Tweets were identified if they contained the words ``eat'' and ``bats.''
    \item \textbf{War Criminal}: tweets that mentioned the words ``war criminal.''
    \item \textbf{Made in China}: tweets mentioning the words ``made in China''. In the context of tweets that discuss COVID-19, this indicates that the tweeter is tying COVID-19 to China.
    \item \textbf{Genocide Complicity}: tweets that mention the words ``complicity'' and ``genocide.'' The use of the word genocide implies that the tweeter is trying to indicate that COVID-19 is being used as a tool to eliminate a specific population and that someone is complicit in it. 
    \item \textbf{Rape} and \textbf{Torture}: tweets that contain the words ``rape'' and ``torture.'' These words are used in a hyperbolic sense to exaggerate the effects of COVID-19.
\end{itemize}
We created a feature that identified whether these specific tropes were referenced in a tweet.

\subsubsection{Sentiment Analysis Features}
The Twitter content that we identified as trolling was generally very negative, and we perceived the tweets as being aggressive and divisive. The psychological research of trolling confirms these observations and indicates that trolls are characterised with ``deceptive and divisive behaviour'' \cite{Sest}, are ``disruptive'' \cite{Hardaker, Lokk} and generally convey negative sentiment. To capture these trends in our troll content labelling and later in our model, we followed the sentiment analysis approach used in \cite{Fornacciari} and \cite{Capistrano}. To identify the sentiment of tweets, we leveraged the VADER sentiment analyzer to automatically identify negative sentiment in the dataset ~\cite{VADER}.  

\subsubsection{User Behaviour}
Taking into consideration the relevance of the user's behaviour in our detection of trolling behaviour \cite{Fornacciari}, \cite{Riquelme}, we used the type of the tweet (original tweet, a retweet, or a reply to a tweet) to characterise a user's Twitter behaviour. We limited the user behaviour only to the type of tweet because other features, like favorites, citations, or number of replies to a tweet, have provided minor gains for trolling content identification \cite{Fornacciari}. In our case, we used the type of tweet to distinguish a trolling behaviour where a combination of retweets and replies to other tweets and other users indicate an attitude of following and engaging in multiple conversations ~\cite{Samory}.

\subsection{Implementing TrollHunter}
The system architecture of \textit{TrollHunter} is shown in Figure 1. \textit{TrollHunter} consists of three processes (data clean-up, model training and selection and model deployment) which interface with three external entities (Twitter, the labeller and the moderator). The main input into \textit{TrollHunter} is raw data collected using Tweepy (a Python wrapper for the Twitter API) \cite{tweepy}. The first process, \textit{data clean-up}, extracts the relevant fields from the raw Twitter data (the tweet content, type of tweet and the hashtags) before it forwards it to the \textit{labeller}. The \textit{labeller} analyzes this data and outputs labels for whether or not each tweet exhibits trolling behaviour. The \textit{labeller}, in our implementation consisted of human trolling experts who manually determined which tweets were suspected to be trolling tweets based on the categorization elaborated in the section above. While our labelling was all manual, it's possible that future iterations of \textit{TrollHunter} could use a labeling process that is crowd-sourced, labelled using other algorithms, or use semi-supervised techniques. The cleaned, labelled data is forwarded as an input into the \textit{model training and selection} process in our trolling content classifier. Once this model is created, it is trained on the labelled data and provided to the \textit{model deployment} process. The \textit{model deployment} takes the trained model and uses it to flag the new tweets coming in from Twitter as either ``trolling'' or ``non-trolling''. The output of the \textit{model deployment} is forwarded to the moderator. The moderator could refer to a Twitter admin, but in general it can refer to anyone who might use the troll tweet flag to take action. 

\begin{figure*}[h]
  \centering
  \includegraphics[width=0.55\linewidth]{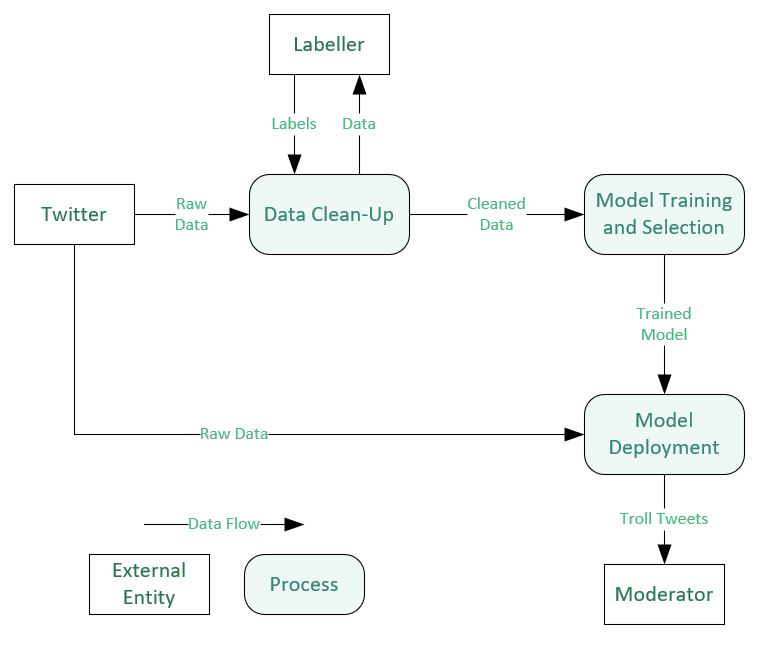}
  \caption{TrollHunter System Architecture}
\end{figure*}

\subsubsection{Possibilities for Unethical Use of \textit{TrollHunter}}
We are aware that there is a potential of unethical use of the \textit{TrollHunter}. In the context of this research, we did not implement any follow-up action (e.g. report, block, suspend account, flag) for the \textit{TrollHunter} to perform as it identifies trolling behaviour. Using any sort of tweet classifier, especially one that does not have any form of human intervention before performing an action, can be abused. In the context of the \textit{TrollHunter}, the tool, with updates to the training data could be used to silence minority opinions, introduce racial bias and hamper Twitter's ability to provide a space for free-speech \cite{Davidson}. This obviously is not the goal of \textit{TrollHunter}. We are imagining \textit{TrollHunter} being used by an Twitter content moderator to help them more efficiently identify trolling content or perhaps identify accounts that consistently exhibit trolling behaviour.

\subsubsection{Data Clean-up}
To clean the tweet text and hashtag attributes, the tweet textual content needed to be converted into numeric vectors before it could go into the \textit{model training and selection} process. To create the vectors, we used a TF-IDF (Term Frequency-Inverse Document Frequency) algorithm. TF-IDF weighs the importance of words to give less importance to words that show up in lots of documents and increase the importance of words that show up more frequently within a specific tweet. This helps to ensure that the model does not consider words that show up in lots of tweets (like ``coronavirus'') as important as words which appear in fewer tweets. We also removed stop words, which are common words that frequently appear in English text like ``the'' or personal pronouns. Finally, we made all text lowercase and removed all words that appear in fewer than 1\% of tweets. The hashtags in each tweet were combined into a single string as a separate column, with the hashtag character removed, then they were cleaned and utilised in exactly the same manner as the tweet text. 

\subsubsection{TrollHunter Model Design Decisions}
In addition to the explicit goal of identifying trolling behaviours, we built the \textit{TrollHunter} in the context that we would be attempting to break it using the TTE paradigm along with our application of a Markov chain to manipulate the text within tweets. Therefore, we elected to make several decisions to make the model more robust to attacks. We made two design decisions which limit the model's ability to generalise well to trolling tweets in other contexts, in the hopes that these additional steps might help reduce the effectiveness of the \textit{TrollHunter-Evader}. First, we decided to perform complicated feature engineering, using features that we identified while labelling the data and re-added the hashtags totally separately from the text. Second, we decided to use a model that used all of the features of the dataset, rather than a subset, because we reasoned that the more words that the model accounted for, the more words which we would need to change in order to make the \textit{TrollHunter} misclassify the trolling tweets and increase the likelihood that a casual viewer could identify the suspicious tweet.

We decided to use a support vector machine, which is a well-established reliable model that draws a line between the different features in a multi-dimensional space based on the optimal line between the trolling and non-trolling tweets \cite{han_svm}. While support vector machines allow the line separating the classes to have one of a variety of different shapes depending on the selected kernel (such as a \texttt{sigmoid}, a Gaussian radial basis function, or  polynomial), we found, through experimentation and cross-validation, that a linear kernel gave the most reliable results. As such, our experiments confirmed that it is the least prone to over-fitting and trained the fastest \cite{han_svm}. For our specific implementation we decided to use the Scikit-Learn implementation of a linear support vector machine, called \texttt{LinearSVC} \cite{scikit-learn}. Our selection of the support vector machine is important because of the high-dimensionality of our dataset. In total, based on the feature sets described in the previous section, our model contained 43,120 features (18 features created in the previous section, as well as 40,045 features from the vectorised tweets and 3,057 features from the vectorised hashtags). 

Other models would have been more susceptible to this type of attack. For example, a decision tree classifier, which only uses a subset of the model's features that make the biggest impact on the decision \cite{Huang}, would have fewer words that the adversary would need to change to affect the model's performance. Therefore, if an adversary were targeting a decision tree, they might only need to change a couple of words to make the tweet a trolling or non-trolling tweet and then they could force the model to misclassify trolling tweets 100\% of the time. For a more thorough discussion of how the \textit{TrollHunter} could be further hardened, see Section 5.3.

\subsection{TrollHunter Performance}
To fine tune the performance of the \textit{TrollHunter}, we focused on improving our recall (the likelihood that our model correctly identified a trolling tweet) and our precision (the likelihood that a trolling tweet was correctly labelled). Because our dataset was imbalanced in respect to the number of trolling (3\% of tweets) and non-trolling tweets (97\%), we used an oversampling technique called ``SMOTE'' to create enough fake tweets and restore the balance \cite{SMOTE}. SMOTE algorithmically creates synthetic samples by generating numerical vectors that mimic the numerical representations of existing vectors which were actually trolling tweets. Without these artificially-generated samples, the model would predict that all tweets are non-trolling which would result in a 97\% accuracy. Table 1 shows the results of a 10-fold performance validation of \textit{TrollHunter}. 

To frame our results more explicitly, in our data set, the model correctly identified 69.8\% of trolls. Of the trolls our model identified, 75.4\% were actually trolls. In total, 98.5\% were correctly categorised (regardless of whether they were trolls or not). Our raw performance is comparable with existing troll detector models. In terms of accuracy, our model's 98.5\% outperformed the troll detector in \cite{Capistrano}, who reported 88.95\% accuracy, but their model was better in terms of precision (they got 86.9\% in comparison to our 75.4\%) and recall (93.1\% to our 69.8\%). We believe that our results are promising given that in our case we created our own dataset focused on an emergent trolling campaign, unlike the test trolling dataset used in \cite{Capistrano}. In other words, the \textit{TrollHunter} model is very specific to our dataset and reliant on the feature engineering we performed to find specific trolling patterns relevant to the COVID-19 infodemic. 

\begin{table}[tbhp]
\renewcommand{\arraystretch}{1.0}
\caption{TrollHunter: Performance} 
\label{table_2}
\centering
    \begin{tabular}{|c|c|} 
         \hline
         \textbf{Metric} & \textbf{Score}  \\
         \hline     
         Accuracy & \(0.985\) \\
         Precision & \(0.754\) \\
         Recall & \(0.698\) \\
         F1 Score & \(0.687\) \\
         \hline
    \end{tabular}
\end{table}

\subsection{TrollHunter: Future Adaptations}
For our initial test of \textit{TrollHunter} we targeted trolling tweets that spread disinformation or unverified rumours about the COVID-19 pandemic. A future adaptation of the \textit{TrollHunter} could hunt for less granular but more informative trends of what Starbird calls ``alternative narratives'' about the COVID-19 pandemic \cite{Starbird}. The alternative narratives on Twitter run counter to the mainstream narrative about polarized topics, spreading viewpoints based on false information, conspiracy theories, and rumors, in series of tweets or within a particular Twitter community. A single tweet adding an implicit reference to an alternative narrative about seemingly benign information might not be detected by the current version of \textit{TrollHunter} but nonetheless helps amplify an alternative narrative about the COVID-19 pandemic on Twitter. For example, a tweet with a benign hashtag \#COVID19 talking about the confirmed re-emergence of the swine flu (H1N1) in 1977 as a result of a lab mistake won't be detected by \textit{TrollHunter} but implicitly refers to the conspiracy theory that the COVID-19 virus originated in a laboratory in Wuhan ~\cite{Sanger}. We plan to implement this adaptation of \textit{TrollHunter} in continuation of our work and test it with Twitter data collected during the 2020 US elections.  

\subsection{TrollHunter: Implications}
Zannettou et al. showed that alternative narratives flow through multiple social media platforms like Reddit, Twitter, and 4chan \cite{Caulfield}. The introduction of automated trolling detection, in general, could potentially have an effect of moving COVID-19 trolls to less regulated platforms. Although there is some evidence that state sponsored trolls persistently disseminate trolling narratives on Twitter \cite{Im}, the presence of an automated detection can nudge individual trolls or users interested in alternative narratives to other platforms. A recent example of such a migration from Twitter to Parler, Rumble and Newsmax was witnessed after Twitter actively labeled and removed false information on the platform during the 2020 US elections \cite{Isaac}. An opposite effect is also possible, where trolls or fringe Web communities disseminating trolling narratives could be attracted on Twitter by exploiting the limitations of \textit{TrollHunter} to precisely distinguish between ``trolls'' and ``non-trolls.'' Aware of the limitation of the feature engineering of \textit{TrollHunter} described in the previous sections, trolls could come with alternative tropes or hashtags, for example, \#ReopenAmericaNow or \#StopTheMadness spreading an anti-quarantine sentiment, that could evade detection. 

An interesting question then arises about the entropy and usefulness of the trolling tweets that are able to evade detection. Before we move on to describe our approach of automated trolling evasion in the next section, it is worth noting that the 24.6\% of false negatives indicate COVID-19 trolling tweets have a non-negligible survival rate. A similar outcome could be observed, we suppose, if any of the other automatic trolling detection methods were used to hunt for COVID-19 trolls because of limitations in labeling or test/training data split, but also the evolution of trolling tactics. The COVID-19 infodemic immediately responds to the daily influx of new information about the COVID-19 virus so the trolls can use new topics or developments much faster than the labeling and testing can take place. In addition, the labeling usually takes a slightly conservative definition of COVID-19 trolls to account for balanced discourse and avoid perception of excessive control and censorship, which in turn, allows some of the trolling content to remain undetected \cite{Fornacciari}. 

A general implication, thus, is that trolling on Twitter has a \textit{parasitic existence} supported by (1) the human ``intolerance of uncertainty'' especially during an event like the COVID-19 pandemic \cite{Mertens}; and (2) the persistent updating, re-purposing, and developing new alternative narratives for trolling on a particular polarizing topic \cite{Im}. Consequently, \textit{TrollHunter} is not able not detect the ``seed'' that makes the COVID-19 topic ``polarized'' on Twitter in the first place -- at least not when hunting for individual tweets. This is another reason for us to adapt \textit{TrollHunter} to hunt for ``trolling narratives'' instead, as we noted above. Certainly, a future research can look how this adapted version of \textit{TrollHunter} could be customized to follow the flow of the COVID-19 narratives on multiple platforms like Reddit or 4chan. 

In any form, \textit{TrollHunter} remains an experimental tool for automated trolling detection on Twitter. A move to a production platform will entail a full Machine Learning (ML) evaluation and possibly redesign using the principles of Software Development LifeCycle (SDLC) suggested in \cite{Shankar}. Even though an official repository of ML attacks is lacking, \textit{TrollHunter} should be tested and refined against known adversarial ML attacks listed in \cite{Shankar} and \cite{Carlini}, as well being subjected to static/dynamic analysis. An audit, as suggested in \cite{Papernot2018}, will benefit \textit{TrollHunter} to detect any anomalies during the labeling or the ongoing troll hunt, given that the COVID-19 or any other future infodemic evolves in a very unpredictable fashion \cite{Vosoughi}. After the deployment, a registration and handling of vulnerabilities like in the Common Vulnerabilities and Exploits (CVE) system will not just alert Twitter, but help other social media platforms or automated troll hunting systems to check their security posture accordingly. Finally, a production-deployed \textit{TrollHunter} must enable incident response and forensics to ascertain the root cause of failure. All of these aspects will turn \textit{TrollHunter} into a robust system with a recognizable salience in deterring COVID-19 or other trolling narratives.  

\section{TrollHunter-Evader: Evading Automated Trolling Detection}
\subsection{Background}
The \textit{TrollHunter-Evader} employs a new evasion paradigm where an existing tweet, either ``trolling'' or ``non-trolling'' can be recycled and repurposed for a persistent trolling campaign without the need to constantly develop new trolling content to evade detection. The paradigm behind \textit{TrollHunter-Evader} builds on two bodies of research. The first is research on \textit{adversarial machine learning}, which is a study of how adversaries can gain an advantage over a deployed machine learning model by figuring out how the model works and changing the model's decision boundary, or its inputs until it mis-classifies the input \cite{Huang}. The second body of research is on \textit{ambient tactical deception} or ATD \cite{nspw2019}, which is a linguistics manipulation attack targeting the textual content of a web page, social media post, or an email with the goal to deceive a target user. 

\subsubsection{Adversarial Machine Learning}
To evade the classifier, we are primarily concerned with developing a variant of the Test-Time Evasion attack (TTE) \cite{Miller}. The TTE attack is an adversarial machine learning method where the adversary aims to avoid detection by manipulating malicious test samples \cite{Biggio}. The goal of the TTE attack is to force a classifier to make a judgement that is not in line with human consensus. The basic TTE adversarial model suggest that the adversary’s capability is limited to modifications of input data, that is, the target machine learning method is assumed to be a black-box. To implement a TTE attack against Amazon Web Services and Google Cloud Prediction, authors in \cite{Papernot} created a so-called ``substitute model'' using an oracle access to the target classifier. They used several different mathematical algorithms to craft the adversarial test samples and showed that these samples are equally effective in misleading both the locally trained substitute model and targeted classifier. In similar regards, \cite{Tsingenopoulos} created a reinforcement learning framework where an adversary can query a black-box target classifier to extract the underlying decision behaviour and accordingly craft adversarial samples using customised algorithms. 

We used the same conceptual approach in creating the \textit{TrollHunter-Evader}, although we employed, in part, the ATD linguistic manipulation on the trolling tweets to craft the adversarial input samples. For the other part of creating the adversarial input samples, we utilised the Markov chain approach \cite{Song}. Authors in \cite{fursov} have already demonstrated that a Markov chain algorithm can be used to select target words for replacement when crafting adversarial input samples from. In \textit{TrollHunter-Evader} we extended the Markov chain approach to rely more heavily on heuristics in order to make subtle textual changes in the adversarial samples that look like they belong to a human, given that our objective is evading trolling detection by the \textit{TrollHunter}.

\subsubsection{Ambient Tactical Deception (ATD)}
The ambient tactical deception as a concept stems from the effort to explore alternative ways of instilling negative sentiment and divisiveness, much like trolls do, through linguistic manipulation of social media content \cite{Sharevski1}. The difference in the ATD method is that the manipulation is not made by directly crafting a trolling post, tweet, or a comment, but instead, it is done by a man-in-the-middle malware that intercepts an original non-trolling content, rearranges the words and text and presents it to a target user to achieve a similar effect to trolling \cite{Hardaker}. The malware is packaged as a browser extension, an email client ``add-in'' (e.g. Outlook), or an entirely new application \cite{nspw2019}. Developing extensions, add-ins and apps is free and a benign software can pass all the security checks before publishing \cite{Newman}. For example, a browser extension variant of the malware can disguise the ambient tactical deception logic and pass the security checks by posing as an ``accessibility (a11y) extension'' that claims the rewording is done to help non-native English speakers make sense of English slang on social media \cite{Jang}. 

The linguistic manipulation of online communication is specific to a target set of individuals (e.g. linguistic style, pragmatics, cultural norms, topics, etc.), therefore, the ATD attack so far has been tested as a method for conveying divisiveness in Facebook debates on freedom of speech \cite{Sharevski1} and Twitter vaccine debates \cite{Sharevski2}. In the Facebook debate, the malware intercepted originally liberal-leaning comments on a Facebook post and made them look conservative-leaning (e.g. the original comment ``It's hard to be a liberal kid at school - the far-right zealotry will eat you alive!'' was made to read ``It's hard to be a conservative kid at school - the far-left zealotry will eat you alive!''). In the Twitter debate, the malware intercepted a pro-vaccine tweet and made it look anti-vaccine (e.g. the original tweet ``No serious academic journal articles support the claim that there is a link between autism and vaccines!!! \#provax \#vaccineswork'' was made to read ``Many serious academic journal articles support the claim that there is a link between autism and vaccines!!! \#antivax \#vaccinesdontwork''). The ATD, in both studies, proved as a potent way of making the social media users perceive a legitimate content as being overtly ``negative and divisive.''

\subsection{Crafting Adversarial Input Samples}
In the initial ATD work, the adversary manually inferred the linguistic manipulation of the content in the context of the social media discourse. This limited a wider scalability of the attack as well as automated adaptations to evade human detection. Another adaptation of the initial ATD work is needed to extend the use of opposite-polar approach (to achieve divisiveness) in order to account for promulgating disinformation. Therefore, we extended the ATD paradigm to allow for automated \textit{context-relevant word substitutions}. We created an Markov chain algorithm for crafting context-relevant word substitutions in a candidate text for an adversarial input sample. In this section, we describe our algorithm using a minimal example from the first verse of the poem “Jabberwocky” from \textit{Through the Looking Glass} by Lewis Carroll \cite{carroll}: \\

\begin{center}
    \begin{verbatim}
        ‘Twas brillig, and the slithy toves
        Did gyre and gimble in the wabe;
        All mimsy were the borogoves,
        And the mome raths outgrabe.
    \end{verbatim}
\end{center}

We divided the stanza into two pieces, a training set and a target adversarial sample set. Using the first two lines, we prepared a Markov chain with an initial and a final state, along with the probability that the initial state will be followed by the final state, as shown in Table 2. Next, we selected target words employing the ATD approach. In our case, we targeted occurrences of the words “borogoves” and “mome”: \\

\begin{center}
    \begin{verbatim}
        All mimsy were the TARGET,
        And the TARGET raths outgrabe.
    \end{verbatim}
\end{center}

\begin{table}[!h]
\renewcommand{\arraystretch}{1.0}
\caption{Markov Chain for Crafting Adversarial Textual Samples: States and Probabilities} 
\label{table_2}
\centering
    \begin{tabular}{|c|c|c|} 
        \hline
         Initial State & Final State & Probability  \\
         \hline     
         ‘Twas & brillig, & 1 \\
         brillig, & and & 1 \\
         and & the & 0.5 \\ 
         the & slithy & 0.5 \\
         slithy & toves & 1 \\
         toves & Did & 1 \\
         Did & gyre & 1 \\
         gyre & and & 1 \\
         and & gimble & 0.5 \\
         gimble & in & 1 \\
         in & the & 1 \\
         the & wabe; & 0.5 \\
         \hline
    \end{tabular}
\end{table}

The word preceding the target word each time is the word “the.” When we look in Table 2, we see that there’s a 50\% chance that the word “the” is followed by the word “slithy” and a 50\% chance that the word “the” is followed by the word “wabe.” We did a weighted random choice using the “probability” vector, transforming the full poem reading to read like this: \\

\begin{center}
    \begin{verbatim}
        ‘Twas brillig, and the slithy toves
        Did gyre and gimble in the wabe;
        All mimsy were the wabe,
        And the slithy raths outgrabe.
    \end{verbatim}
\end{center}

While this does not fit the poetic structures in place, someone who is not familiar with “Jabberwocky” might not notice the changes and the nonsensical vocabulary looks like it belongs. Running the ATD on the same passage of text using the same input text multiple times might yield different results, however, the poem still has a similar cadence:\\

\begin{center}
    \begin{verbatim}
        ‘Twas brillig, and the slithy toves
        Did gyre and gimble in the wabe;
        All mimsy were the slithy,
        And the slithy raths outgrabe.
    \end{verbatim}
\end{center}

To leverage the transferability of this algorithm to a dataset of tweets, we created a substitute model of \textit{TrollHunter} that only uses text features and looked at what words help it to make its prediction, then use a Markov chain to replace trolling words with contextual replacements. Due to practical constraints and our visibility into the \textit{TrollHunter} algorithm, we used the same labelled dataset and a similar pipeline to the \textit{TrollHunter} model. This is unlikely to have a major effect on the performance versus the performance that we would have if the targeted classifier were black-box \cite{Papernot}. 

\subsection{Crafting Trolling Evasion Tweet Samples}
A component of the test time evasion paradigm that the \textit{TrollHunter-Evader} relies on is that once an adversary has a local model, they can use a few different techniques to see what the model is ``thinking,'' and remove words from the text which are correlated with the tweet being identified as a trolling tweet. In this subsection, we go in depth into how we applied this paradigm to a specific tweet. 

\subsubsection{Building the Local Model and Finding Target Words}
The adversary does not necessarily know what type of classifier the \textit{TrollHunter} is using, or what sort of feature engineering the model has. Fortunately, the only thing the adversary needs is a general idea of what sort of question the \textit{TrollHunter} model solves, which is that it identifies trolling content on Twitter on the topic of COVID-19. First, the adversary needs to build a dataset. The adversary would most likely have tweets that they are trying to disguise, recycle, or repurpose, which they could compare to non-trolling tweets pulled directly from Twitter. While the \textit{TrollHunter} has to \textit{assume} that the dataset was accurately labelled, the adversary \textit{knows} their dataset is labelled correctly. Once the adversary has a dataset, which they can also balance by themselves, the next step is to determine what features they want to use in their local model.

To determine a list of target words (and hashtags) from the local model, we used a two-step process to, first, understand the impact of each word on the local model's prediction, then to ensure that the values make the classifier determine that a tweet is a trolling tweet. Using the \texttt{scikit-learn} algorithm, the adversary can utilise the \texttt{.features\_importance\_}, which shows the relative importance of each feature in making the final prediction \cite{scikit-learn}. Feature importance leverages the Gini impurity index to decide the importance of each attribute \cite{scikit-learn}. The Gini impurity index, $Gini(D)=1-\sum_{i=1}^m p_i^2$ shows the probability, $p_i$, that a tweet in $D$ is a trolling tweet, calculated across the total number of classes, $m$ \cite{han_dt}. The feature importance takes all of the Gini impurity indexes from each of the features and uses the index to determine how big of an impact each word has on the classifier's prediction with 0 indicating no importance and higher values indicating more importance to the prediction. The sum of all of the feature importances is 1. Because each feature represents a word, we were able to use the feature importances to create a vector showing whether each tweet contained the word. We calculated the Pearson's correlation using the SciPy library and added any word which had a positive correlation with trolling labels and a p-value less than .05 to the list of target words. Positive correlations show that the words are associated with trolling tweets. Removing those words from tweets means that the local model will not identify the tweet as being a trolling tweet. All of the words identified as having a positive correlation with trolling tweets can be used as target words for the ATD engine. 

\subsubsection{The ATD Engine}
The ATD engine uses a Markov chain to rewrite text to \textit{replace} the target words. The Markov chain was trained using 400,000 tweets. The ATD engine can use both regular words and hashtags as a replacement in order to maximise the likelihood for evasion. An example of the ATD engine output is shown in Figure 2. The original tweet in Figure 3 was recycled by replacing the word ``coronavirus'' with the hashtag ``\#COVID19,'' replacing the hashtag ``\#trumpvirus'' with ``\#coronavirus'' and removing the ``\#chinesevirus'' hashtag. Unlike in the limited demonstration of the algorithm in the previous sample, it might not always be feasible to manually select words that the Markov chain could find replacements for.

\begin{figure}[!h]
  \centering
  \includegraphics[width=0.75\linewidth]{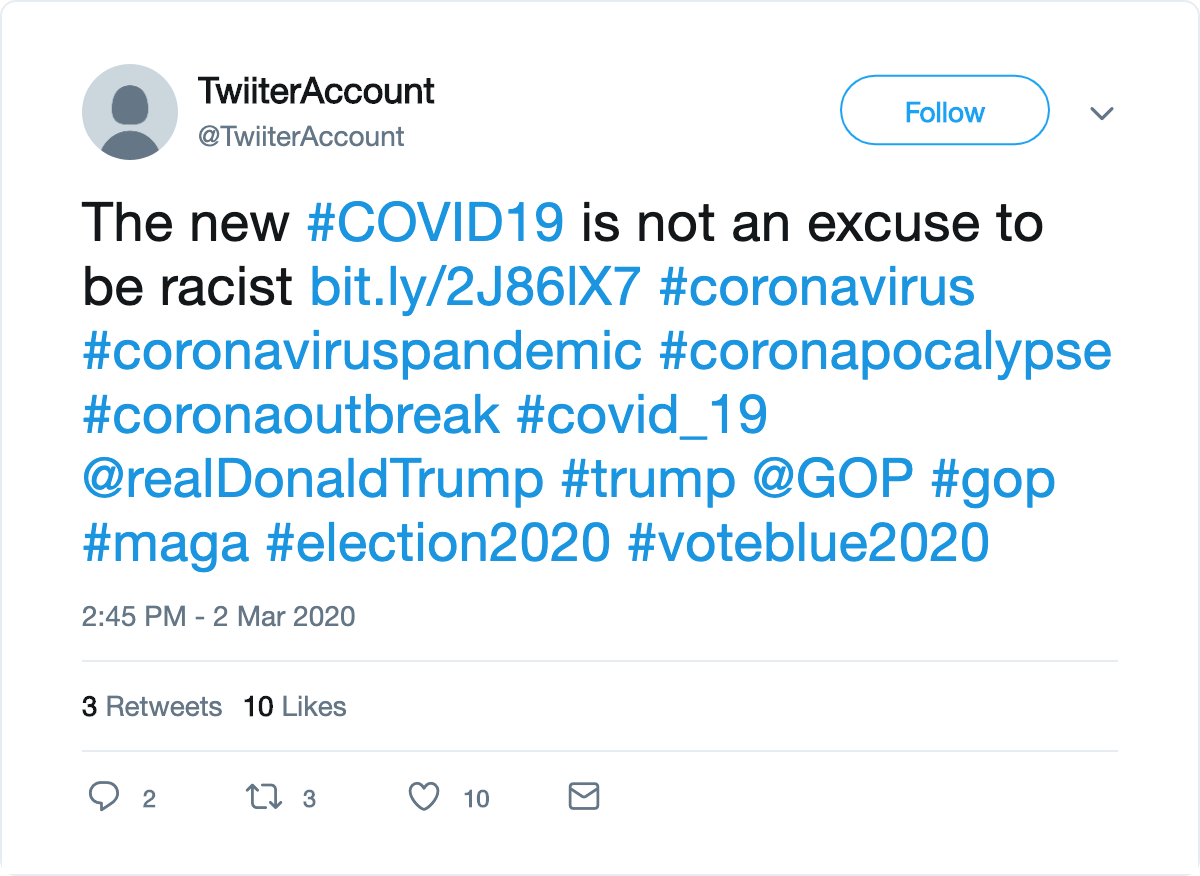}
  \caption{Trolling Evasion Tweet}
\end{figure}

\begin{figure}[!h]
  \centering
  \includegraphics[width=0.75\linewidth]{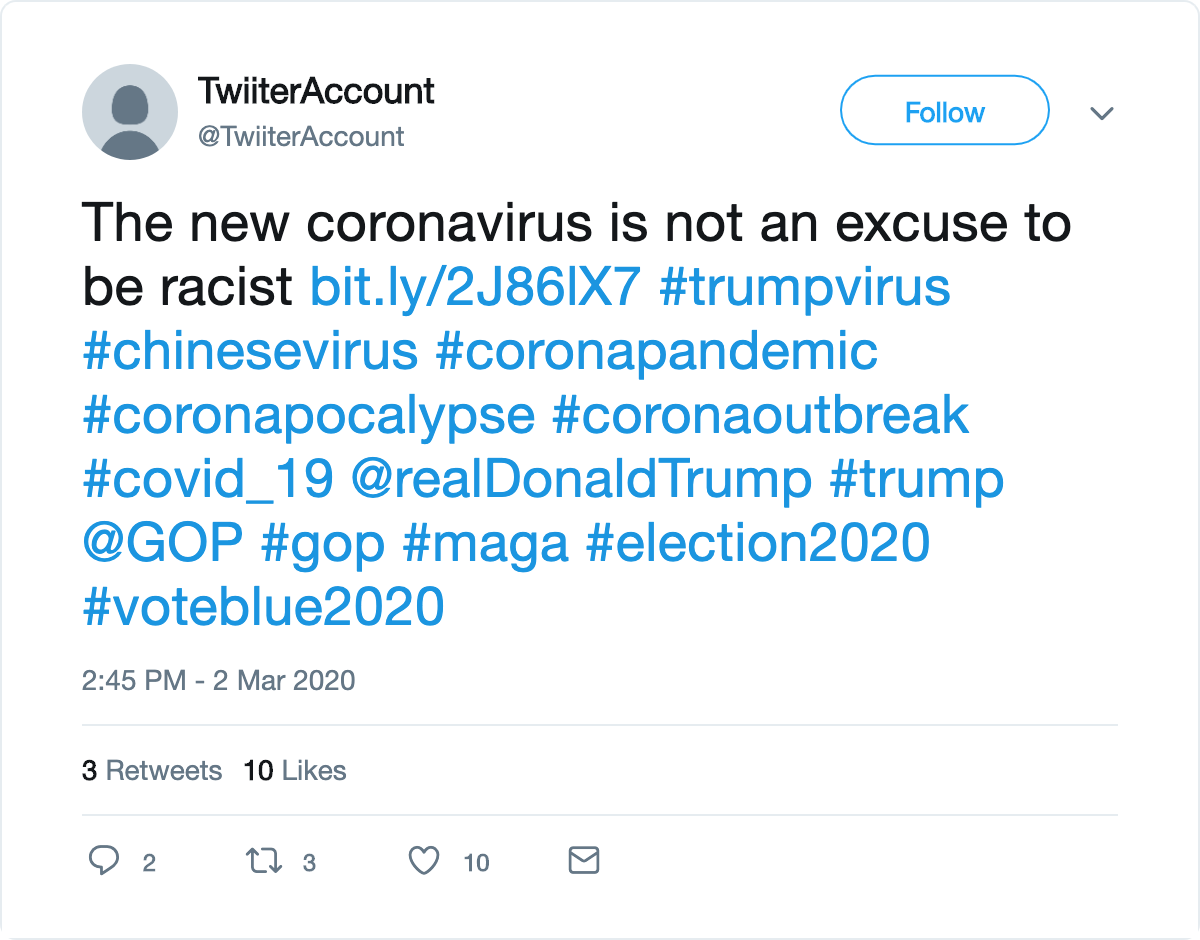}
  \caption{Original Tweet}
\end{figure}

Due to the richness of Twitter vocabulary that can include misspellings, links, hashtags, abbreviations, emojis, borrowed non-English words and slang on top of all other English vocabulary, the Markov chain could not always provide a recommended replacement for a given initial state. In these situations, the Markov chain simply removed the word. This was a case with long strings of hashtags, where a hashtag might be associated with trolling tweets and the preceding hashtag might not have been referenced in the 400,000 tweets represented in the Markov chain, so the hashtag would just be removed. This generally does not affect the message of the tweet, though it has the possibility to adversely affect the coherency of the tweets (which actually comes along with the native trolling behaviour \cite{Hardaker}. 

\subsection{Implementing TrollHunter-Evader}
Based on the paradigm elaborated above, the system architecture of \textit{TrollHunter-Evader} is shown in Figure 4. To evade trolling detection, \textit{TrollHunter-Evader} uses a combination of a local model and an ATD engine that relies on a Markov chain to suggest context-relevant word substitutions in the tweets. The Markov chain essentially is created from a substantial corpus of tweets that is assessed, word by word (hashtag), to determine whether each word is a target word (hashtag). If a word in the text is a target word, the ATD engine assigns a new word based on the preceding words (If the preceding words are not available, the target word is removed). Using \textit{TrollHunter-Evader} to successfully evade a classifier like the \textit{TrollHunter} requires two things: first, a strong list of target words that the ATD engine will overwrite; second, a corpus that contains similar enough text to what an adversary wishes to corrupt. 

\begin{figure*}[h]
  \centering
  \includegraphics[width=0.755\linewidth]{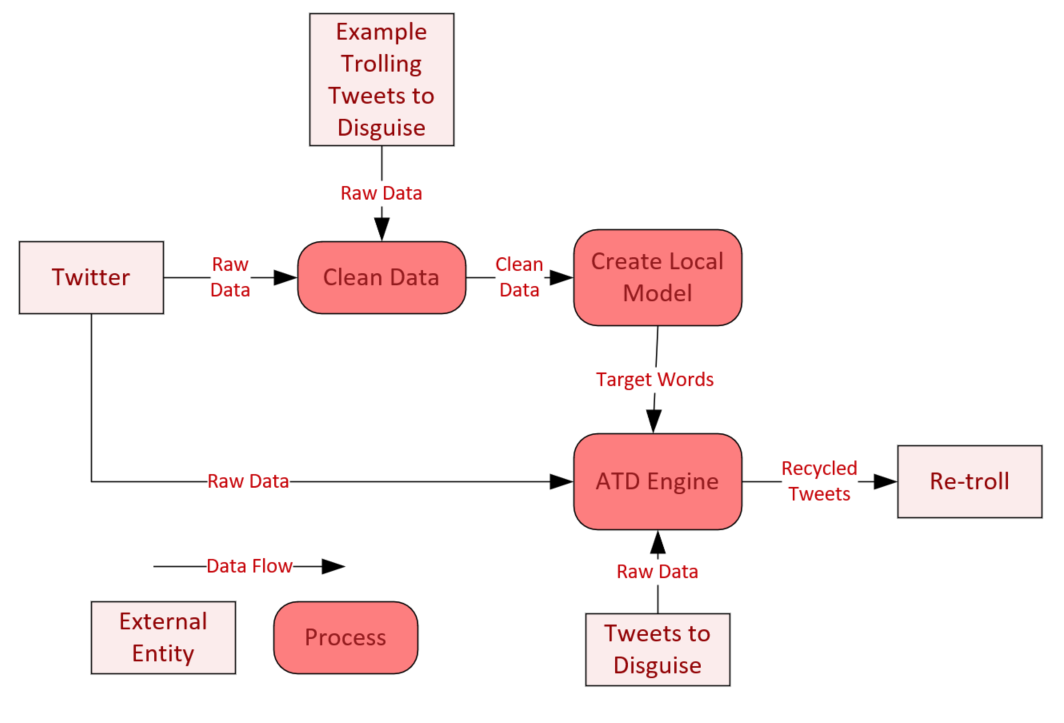}
  \caption{TrollHunter-Evader System Architecture}
\end{figure*}

Although our strategy for \textit{TrollHunter-Evader} was to beat a classifier using a test-time evasion technique using target words that we derived from a local model, in practice, an adversary might use a different strategy. So far we have experimented with three different strategies for identifying target words. These are to pick, based on context, what an adversary wishes to replace manually in order to evade detection but still achieve a trolling outcome. This is effective in cases where we wish to censor a specific piece of information. For example, an adversary might try to publish a version of the Mueller report that does not reference ``Facebook'' or ``Mark Zuckerberg.'' They would simply select those words and let the ATD engine make contextual replacements each time one of those words appeared. The next strategy is to randomly make replacements based on arbitrary rules, e.g. replacing 10\% of verbs. If the adversary has a Markov chain trained on articles discussing corruption, that could help generally make the subject of a piece seem more corrupt. Finally, the strategy that we used for \textit{TrollHunter-Evader} is based on the feature importance calculated based on the local model. We also experimented with some success by using the permutation importance deived from a local support vector machine.

\subsection{TrollHunter-Evader Performance}
To test the effectiveness of the \textit{TrollHunter-Evader} on making the original \textit{TrollHunter} misclassify trolling tweets, we ran a preliminary test on a set of 100,000 new tweets. We used \textit{TrollHunter-Evader} to corrupt any tweets that were identified as exhibiting trolling behaviour by \textit{TrollHunter} and re-ran the tweets through \textit{TrollHunter} again. In total, \textit{TrollHunter-Evader} forced the \textit{TrollHunter} to misidentify approximately 40\% of the tweets. The 40\% success rate is interesting considering that we are targeting specifically the \textit{TrollHunter} with recycled tweets, which in general, significantly decreased the overall performance as shown in Table 3. The reduction of the model's recall to 0.419 means that the \textit{TrollHunter} is now only able to successfully identify 41.9\% of trolling tweets and 57.9\% of the trolling tweets are not identified as ``trolling''. An individual or a state-sponsored troll farm, in this way, could use the \textit{TrollHunter-Evader} output to reduce faith in the \textit{TrollHunter} model and make it so that it's recommendations are no longer taken seriously. 

\begin{table}[h]
\caption{TrollHunter: Performance on adversarial textual samples created by TrollHunter-Evader} 
\label{table_3}
\centering
    \begin{tabular}{|c|c|c|} 
         \hline
         \textbf{Metric} & \textbf{Score} & \textbf{Change in Performance}  \\
         \hline     
         Accuracy & \(0.969\) &\(-0.016\)\\
         Precision & \(0.454\) &\(-0.301\)\\
         Recall & \(0.419\) & \(-0.279\)\\
         F1 Score & \(0.687\)&\(-0.289\) \\
         \hline
    \end{tabular}
\end{table} 

While the \textit{TrollHunter} was comparable to many other trolling detectors, after the \textit{TrollHunter-Evader} disguised the tweets, the \textit{TrollHunter}'s performance was reduced to a no-longer acceptable quality. After ATD was applied to the model, about 40\% of the trolling tweets were no longer correctly identified as trolling tweets by the \textit{TrollHunter}. This means that in total, while the \textit{TrollHunter} correctly identified 69.8\% of trolling tweets, the \textit{TrollHunter-Evader} reduced the performance by 30.1\% down to only correctly identifying 45.4\% of trolling tweets. Of the trolling tweets that the \textit{TrollHunter} identified as troll tweets, 75.4\% of them exhibited trolling tweets, after the \textit{TrollHunter-Evader} performed ATD, the performance was reduced by 30.1\% down to 45.4\%. While the accuracy stayed very high, at 96.9\%, due to the comparatively small number of trolling tweets relative to non-trolling tweets, perfect performance by the \textit{TrollHunter-Evader} would only reduce the accuracy of the model down to 94.5\%.

\subsection{TrollHunter-Evader: Implications}
This approach takes existing established attacks intended to evade classifiers to identify which words in a text sample might lead the classifier to misclassify the text and it uses an automated tool created to increase the speed an adversary could perform ambient tactical deception. This has huge implications that could dramatically increase the effectiveness of troll farms and could increase the influence that they could have on future political environments, or to permanently alter the ability of other troll-detectors to make their predictions. In this section we explore possibilities for some of the mischief that a bad actor could possibly accomplish using the techniques outlined in this paper. 

\subsubsection{Reducing Faith in Troll Warnings}
Someone who is a troll is not inherently bound to only posting trolling content. Using the techniques described in this paper, a troll could take non-troll tweets, then use the model described here to make a reasonable troll detector model flag them as troll tweets. If a moderator tries to ban their account using a classification model, they could appeal the decision and when a human examines the content of their tweet, it will not seem like trolling content, so they will reverse their decision. If enough trolls were to coordinate and do this, they could quickly overwhelm any moderators so they decide not to take future warnings as seriously.

\subsubsection{Two Birds with One Scone}
When combined with ambient tactical deception, for example by changing text to seem less polite to give the target a negative outlook on how they are perceived \cite{nspw2019}, the adversary simply needs to pick words that they would like to have be censored from the text to which they expose their target. Combining these two attacks would mean that, in addition to increasing the difficulty for detecting this content using automated means, the text could be easily re-purposed, reducing the need for expensive human resources to perform an disinformation campaign. 

\subsubsection{Poisoning/Boiled Frog Attacks}
An on-line classification model is a classification model that updates itself with new samples as the model evaluates them. These models, when their input is not validated, e.g. it is pulled directly from Twitter, are susceptible to poisoning and boiled frog attacks, which involve the adversary feeding lots of specifically selected data into the model to change how the model makes its decisions \cite{Biggio1}, \cite{chio}. By continuously feeding trolling samples into a model, an adversary could permanently alter the classification model to consistently misclassify trolling tweets, making it easier for any trolling content to be classified as ``non trolling,'' eliminating the model's effectiveness, allowing any bad actor with a Twitter account to post any trolling content without it being detected by the classifier.

\subsubsection{Honing Social Engineering Attacks}
Using the automated ATD approach of rewriting text allows someone with limited knowledge of English to make automated changes to their text to evade classifiers that detect trolling behaviours. Assuming an adversary has a template phishing email along with a list of email addresses of people who speak that language, they could utilise a tool like the \textit{TrollHunter-Evader} to quickly make the template email less likely to be classified as ``spam'' by a spam classifier. They will likely have other template phishing emails, so all they would have to do is use the local model and the ATD engine along with a corpus of emails that are not spam (e.g. the Clinton corpus \cite{DeFelice}, or they could even create one from their own inbox). 

\section{Discussion} 

\subsection {Limitations to Our Approach}
The hunt for trolls on Twitter during the COVID-19 pandemic using the \textit{TrollHunter} automated reasoning mechanism has several limitations. There were considerable uncertainties and a lack of structured response regarding the COVID-19 in the early stages (January-March 2020), during which time we collected our Twitter data set. Because of this, there is a possibility that the initial process of labelling trolling content, e.g. the trolling hashtags, topics and accounts, might not represent the entire trolling effort that was present on Twitter, both from individual and state-sponsored trolls. Abundant new disinformation and rumors have been promulgated about the COVID-19 since the first quarter in 2020 and it is plausible that the trolls contextually capitalised on this content to their objective. Therefore, the \textit{TrollHunter}'s performance will decline due to changes in the political environment, as well as shifts in trolling tactics. It is also plausible that \textit{TrollHunter} performance might considerably vary with a different selection of training subset of tweets, even though our analysis with an alternative subset of our choice showed no significant difference. Even though \textit{TrollHunter} shows promising results (limited to the particular choice of sentiment analysis libraries and the choice of support vector machine as a classifier algorithm), they are far from detecting the entire content behind the COVID-19 infodemic. Therefore, we caution the use of the features outlined in this paper as a sole decision-maker in regards trolling/non-trolling content, but they could be used in future models in assisting moderators for assessments of identifying some possible trolling tweets. 

\textit{TrollHunter-Evader} also comes with a set of limitations. The evasion performance presented in this paper comes as a result of the particular way we constructed the local model as well as the implementation of the ATD engine. A selection of other TTE approaches than ours might result into a different output of targeted words and hashtags than the one produced by \textit{TrollHunter-Evader}. This difference will consequently impact the output of the ATD engine and might result in a lower yield of recycled tweets, possibly with marginal benefit compared to manual recycling of the tweeter content. In other words, it is plausible that \textit{TrollHunter-Evader} might not be effective in decreasing the performance of \textit{TrollHunter} in a meaningful way. We must mention that even though our analysis indeed showed a degradation of \textit{TrollHunter} by supplying the recycled tweets of \textit{TrollHunter-Evader}, not all the recycled tweets yielded as useful. Roughly 10\% of the recycled tweets in our analysis, after a manual inspection by a human judge, showed no logical consistency in the syntactical structure of the content and as such had to be removed. 

\subsection{Inherent Limitations in Trolling Content Detection} 
Without the potential for perfection, application of a trolling detector at scale has limited utility as a content moderation strategy. Using multiple advantages that a trolling content classifier created by an actual social media company would not have, namely the manual identification of trolling tropes and hashtags, and a full month of analysis, and trial and error that went into creating the \textit{TrollHunter} we built a highly effective classifier that performs in just a narrow window of time. If the rate of trolling tweets we observed, and our model performance remained steady and was applied to 1,000,000 more tweets, there would be about 30,000 trolling tweets, of which, 20,940 would be correctly identified, meaning that 9,060 trolls would not have been identified, and their content would remain online, while 6,832 non-trolling tweets would be misclassified as trolling tweets. 

Because the trolling tweets are identified using a model that is based on human identification, the 6,832 tweets that are incorrectly identified may appear to be trolling to the labelers due to their personal biases. Further, another audience might consider the 3\% of tweets that the model identified as trolling to be totally acceptable, and might feel that another 3\% of tweets are subversive and dangerous. Even if the labelers are on the same page as the users, they may not fully understand the additional implications or cultural context of something that someone tweets. For example, if someone were creating a classifier to reduce the presence of racist content on Twitter, if someone quotes David Duke (a well-known white supremacist) saying ``I love TCP/IP'' the quote itself might not be racist, and the data labelers might not consider the tweet to be racist, but the context that it came from is. This could be perceived by people as if it were, and this may also apply in the context of trolling content. 

Another limiting factor in the detection, or better said, in recognizing trolling content, is Twitter's optimization for user attention and maximum scroll dwell time. Using recommendation algorithms to deliver personalized content to users, Twitter allows users to isolate themselves from viewpoints with which they disagree \cite{Potts}. This leads to ``homopily'' or formation of so-called echo chambers usually centered around the opposite poles of a controversial issue \cite{Garimella}. This sustained, selective exposure to tweets matching users' position could affect how users, as well as labelers, reflect and recognize trolling when they are exposed to a tweet with an opposite stance \cite{Calvillo}. As a result of an illusory truth effect, non‐reflectiveness, or reflexive open-mindedness \cite{Pennycook}, it is entirely possible that users from different echo chambers may have conflicting perception of trolling, forcing \textit{TrollHunter} to split the detection to ``your trolls'' and ``our trolls'' to avoid selection of a biased training set \cite{Wiegand} or being perceived as non-inclusive. 

A useful addition to \textit{TrollHunter} in this context is the usable security recommendations from \cite{Jono}. A demonstration of how COVID-19 trolling is done, why it is done, and the damage of sharing, retweeting, or liking such content could help resolve this conflicting recognition of trolling. In this direction, Twitter regularly updates a COVID-19 misinformation page \cite{TwitterCovid} and independent researchers debunk political COVID-19 information operations in near real-time \cite{Strick}. Another positive suggestion would be full transparency regarding the \textit{TrollHunter} output to the Twitter community. Similar in fashion as the information operations disclosures by Twitter \cite{Twitter}, a dataset, with an equal focus on the COVID-19 trolling content as well as on the trolling accounts, could be posted, regularly updated, and made available to other researchers interested in automatic trolling detection, adversarial machine learning, content analysis, and the socio-political aspects of the COVID-19 infodemic.

\subsection{Troll-Hunter Evader Technical Enhancements}
The evasion paradigm behind \textit{TrollHunter-Evader} is novel and the underlying approach of text manipulation for adversarial purposes could be enhanced in several ways. Some of them, outlined below, we plan to implement in the future. In order to provide a better analysis of the automated trolling detection evasion not just on the COVID-19 topic, we hope to explore and build defenses against these techniques, based on future polarizing discussions on Twitter. In the future, these techniques can be used to protect against future disinformation campaigns that can be used to hijack elections or lead to bad decision-making that could put people in danger (e.g. in the context of misinformation during a pandemic).

\subsubsection{ATD Engine Enhancements}
One of the easiest ways to improve the ATD engine is to incorporate more sophisticated natural language processing techniques. For example, using a part-of-speech tagger, an adversary could ensure that word replacements are grammatically correct by ensuring that the part of speech of the replacement word is the same as the target word. This could be supplemented with a more careful handling of punctuation to ensure that no punctuation is added or removed with ATD recommendations as well as to ensure proper replacement of hashtags. Another possibility is to use a more sophisticated engine for the text-to-text generation of the replacement words, possibly incorporating a deep learning approach into the model.

\subsubsection{Computational Enhancements}
We tested the performance of \textit{TrollHunter} and \textit{TrollHunter-Evader} on a machine equipped with a Intel\textsuperscript{\textregistered} Xeon\textsuperscript{\textregistered} XPU E0165v2 @3.50GHz processor, 16 GB of DDR3 memory and 900 GB of storage space. An adversary with better computational resources could afford to collect a much larger corpus of tweets and with that to produce a larger set of target words and hashtags. The way the ATD engine is set up will allow for increase in the likelihood that a given text replacement would be successful and consequently, increase \textit{TrollHunter-Evader's} ability to deceive machine learning detection models as well as humans. One approach for computational enhancement is using parallel computing to rewrite tweets. This would make it possible to, for example, build a streaming ATD engine able to capture the latest text data and help the local model to output recycled tweets more relevant with what's trending on Twitter in real time. 

\subsubsection{Data Enhancements}
Beyond increasing the size of the corpus, an adversary could consider experimenting with different methods for cleaning the data, for example by correcting the spelling. This could help to reduce the size of the Markov chain by eliminating rows of data that include misspelled words that are unlikely to appear in unseen text (or could be corrected in the unseen text). A similar approach could be done with a thematic grouping of the hashtags corresponding to the most popular COVID-19 pandemic or other contested topics on Twitter (again, there is danger in misidentifying trolling comments in polarised important discussions about important topics that could have serious, unintended consequences outside the social media ecosystem \cite{Lai}). 

\subsection {TrollHunter Counter-Evasion}
The simplest way for a classifier to mitigate the risk of falling victim to the test-time evasion attack implemented by \textit{TrollHunter-Evader} is to rely more heavily on non-text-based features. The \textit{TrollHunter-Evader} only modifies the text and hashtags of the tweet, a classifier that relied on other features, like the number of followers, the Twitter users' description, or the amount of time since they joined Twitter,  because a troll's trolling behaviour occurs in the content of their tweets. While the other attributes are ultimately easy to beat (by creating fake accounts they can use to follow themselves, or to create accounts and not tweeting until their account ages). In other words, trolling happens in text, so as trolling techniques evolve, trolling detection models will need to rely more heavily on text features. The attack is based on modifying a couple of words in each tweet. While changing a word or two in a tweet may seem like a minor modification, tweets are -by definition- pretty short. In our example outlined in Figure 2 and Figure 3, the model changed two words in a 24 word tweet, indicating that those two word swaps accounted for a modification of 8.3\% of the total information feeding into the model. 

\textit{TrollHunter} was not completely susceptible to the evasion (the detection performance decreased by 40\% rather than 100\%). This is in part because we took a couple of precautions to make the model less susceptible to text manipulations (also used by the ATD paradigm). We theorised that the independent variables created during the feature engineering added small additional weights to certain words and hashtags that would reduce the likelihood that the attacker's local model would be able to fully account for the additional weight given to certain words based on the derived features. Adding trolling hashtags manually, it was more difficult for the local model used to develop the target words to mimic the \textit{TrollHunter}, hampering the ATD engine's ability to evade detection. 

An adversary could make a point of using target words that common sentiment analyzers use to identify emotions. To help protect against this potential, defenders can increase the number of words used to determine the sentiment of trolling tweets, for example customizing the sentiment analyzer by adding additional context-specific words, for example related specifically to the COVID-19 pandemic (e.g. the word ``bat'', which is not negative in of itself, but in our dataset, it is associated with a negative stereotype). Additional resilience can be built into a classification model by making sure that a balance of features are leveraged and possibly by adding additional weight to features that must be updated manually. 

Ultimately, regardless of whether the attack is successful is not just whether the classifier misclassifies the tweet, but it also depends on whether the tweet is coherent and doesn't look out of place to the tweet's reader. To do that, the classifier ideally would require as many word substitutions as possible to trigger a reader to disregard the text. To force the classifier to modify more words, we leveraged a support-vector machine classifier that utilises all of the features in the dataset to help make a distinction between a ``trolling'' and a ``non-trolling'' tweet. This meant that the total number of words that needed to be manipulated were greater, increasing the complexity of ATD and increasing the likelihood that an end user would identify that the tweet is not organic.

To force the \textit{TrollHunter-Evader} to modify more words, we could have created adversarial samples, a strategy that has been advanced in multiple papers like ART (the Adversarial Robustness Toolbox) \cite{Nikolae} or CleverHans \cite{Papernot1}. This  strategy is more complicated when working with text data than in other domains, because unlike other use cases, for example in computer vision, where a few pixels in an image could be modified  slightly, for example, by modifying the shade of blue used in a specific pixel in an image to a degree that's not visible to a human, working with discrete variables like short text samples is comparatively more challenging \cite{Morris}. The word ``coronavirus'' can appear in a tweet one time, or two times, but it can't  appear exactly one and a half times. In the context of the \textit{TrollHunter-Evader}, this is not something that an attacker can do. In future we intend to use the \textit{TextAttack} \cite{Morris} toolkit to augment the training data by creating adversarial samples to harden the model against the \textit{TrollHunter-Evader}.

\subsection {Ethical Implications}
We openly acknowledge that both \textit{TrollHunter} and \textit{TrollHunter-Evader} could easily be abused in multiple contexts for nefarious purposes with malicious re-purposing of the Twitter trolling definition provided in Section II. In particular, modification to the labels could be used to isolate people on Twitter who hold specific viewpoints and could be used to identify targets for harassment. For example, it could be used to target Black Lives Matter supporters and then be used with an automated bot to post harassing comments to silence them. Employers could also abuse this approach to affect hiring decisions, or it could be used in countries where free-speech is not a right to identify and punish dissenters. Our detection evasion algorithm can also easily be modified to allow adverse content (like hate speech, racism, or violent language as defined in the Twitter or Facebook inappropriate content policies \cite{hst}, \cite{hsf}) that might automatically be detected to linger longer than what may otherwise be possible until manually detected. We believe that the risks of a potential adversary reading of our research and appropriating this approach with malicious intent is comparatively small in comparison to the risk that an adversary may develop similar capabilities independently to cause damage to online communities. Therefore, we hope that by discussing these techniques, we can shift discussions towards thwarting future attacks.

We believe that when automated reasoning mechanisms take on cognitive work with social dimensions - cognitive tasks previously performed by humans - the mechanisms inherit the social requirements \cite{Frankish}. The \textit{transparency} of the automated reasoning mechanisms is important because: 1) it builds trust in the system by providing a simple way for users and the wider society to understand what the mechanism is doing and why; and 2) exposes the mechanism’s processes for independent revision \cite{Bryson}. In our case, the main function of the \textit{TrollHunter} mechanism is to hunt for trolling content on Twitter during the COVID-19 pandemic. Like most trolling detection mechanisms, \textit{TrollHunter} is far from perfect and incorporates a level of human subjectivity that affects the initial labelling of trolling tweets. Therefore, we developed the \textit{TrollHunter} architecture to be generic enough to allow for use of different trolling labelling criteria, sentiment analysis approaches and user behaviour modeling. We also ``exposed'' or presented the \textit{TrollHunter} system architecture to a sufficient degree that allows for independent revision. For interested parties, we are open to share more details about \textit{TrollHunter}.

The ability to correctly detect 69.8\% of the trolling tweets leaves a considerable room for content mislabelling. Given that the social media platforms aim for a high degree of desired participation and constructive public discourse, a frequent mislabelling that will result in a suspension of a Twitter account and/or removal of content might create a perception of excessive control and censorship \cite{Fornacciari}. On the other side, the social media platforms are essentially \textit{responsible} for addressing the COVID-19 infodemic for which automated trolling detection models are necessary tools aiding moderators' hunt for trolls. The responsibility, in part, then falls on the automated trolling behaviour detection models like \textit{TrollHunter}. This entails a commitment for continuous improvement of the \textit{TrollHunter} performance, which is something we are dedicated to and already working on. 

Similarly, the main function of the \textit{TrollHunter-Evader} mechanism is to recycle tweets in order to evade detection. This, in essence, is a nefarious goal that provides advantage to adversaries interested in continuing the COVID-19 infodemic, which could have adversarial effects on the response to the COVID-19 pandemic. However, we believe that by transparently sharing the paradigm and the structure of \textit{TrollHunter-Evader} the adversarial advantage can be exposed and with that, ultimately become a trolling detection advantage because the troll hunters and moderators will have the knowledge of this evasion approach. Even more, the ``exposure'' of \textit{TrollHunter-Evader's} system architecture allows for consideration of crafting adversarial test samples for other types of automated reasoning mechanisms, outside of \textit{TrollHunter}, that aim to detect abnormalities in any textual content. We utilised the recycled trolling content from \textit{TrollHunter-Evader} only in local tests with \textit{TrollHunter} and never posted or disclosed details of the content itself outside of the one example in this paper. In the context of moving the analysis towards full-scale tests of \textit{TrollHunter-Evader}, a \textit{responsible disclosure} would entail contacting Twitter and working through the details of the detection/evasion paradigms and the structure of the recycled tweets.

In our research, we use a limited dataset that accurately predicts general trolling tweets during the early stages of the COVID-19. Our model should not be used outside of this narrow scope. For example, the tweet dataset was collected before the Centers for Disease Control (CDC) started recommending the wearing of masks as a measure to curb the rapid rise in new COVID-19 cases. As a result, the model would not be able to identify tweets intended to troll readers for wearing masks (or not wearing masks). By extension, \textit{TrollHunter}, if used in a context outside of the pandemic, would obviously fail at successfully identifying trolling content. We also strongly discourage the automated application of classifiers to make decisions about people's ability to have a voice in online communities. 

Although the context in which we implemented \textit{TrollHunter} was in a lab and was studied to measure the impact of our Test-Time Evasion model, the model advanced in this paper may disproportionately mis-attribute trolling intention to tweets written by minorities and people who speak English as a second-language. Davidson, Bhattacharya and Weber identified a similar phenomenon in datasets used to identify hate speech, noting that the datasets were more likely to label tweets written in African-American English as being hate-speech significantly more frequently than tweets written in Standard English \cite{Davidson}. These patterns, due to the biased labelling affect a classifier's ability to make non-biased judgements on whether a text-sample contained hate speech, possibly leading to additional harm to communities which the classifier is intended to help \cite{Davidson}. According to researchers Vidgen and Derczynski, there are a couple strategies that we could apply in future iterations of our research to mitigate these results, such as building multiple binary classifiers to identify specific types of trolling (e.g. Sinophobic COVID-19 misinformation) rather than all misinformation. Re-framing the problem like this could help to reduce the bias present in the model before moderators implement the model in contexts that could lead to the model making decisions that adversely affect the user (for example flagging a tweet as trolling or deleting the tweet) \cite{Vidgen}. 

\section {Conclusion}
In this work, we introduced two automated reasoning mechanisms for detecting and evading trolling detection we named \textit{TrollHunter} and \textit{TrollHunter-Evader}, respectively. \textit{TrollHunter's} goal is to hunt for trolls during the COVID-19 pandemic on Twitter. In  performance analysis of over a dataset of 1.3 million, \textit{TrollHunter} was able to correctly detect trolling tweets 69.8\% of the time. To investigate the possibility of trolling detection evasion, we developed \textit{TrollHunter-Evader}, who's goal is to recycle trolling tweets in order to trick \textit{TrollHunter} to misclassify them as ``non-trolling.'' By manipulating the text and the hashtags in the tweets, \textit{TrollHunter-Evader} was able to decrease the \textit{TrollHunter} detection performance for a remarkable 40\%. We hope our results inform the security community about the potential of automatic trolling detection and the implications of adversarial machine learning evasion. Our goal with this research is to provide an avenue for further analysis of the automatic detection/evasion approach in a broader societal context, considering the inherent threats of algorithmic bias and data labelling subjectivity, especially during uncertain times such as during the COVID-19 pandemic or a nationwide civil unrest.  

\bibliographystyle{ACM-Reference-Format}
\bibliography{covid-19}

\end{document}